\documentclass[%
 reprint,
 showkeys,
 amsmath,amssymb,
 aps, prb,
longbibliography
]{revtex4-2}

\usepackage{graphicx}
\usepackage{dcolumn}
\usepackage{bm}
\usepackage[colorlinks=true,
            citecolor=blue,
            linkcolor=blue,
            urlcolor=blue]{hyperref}

\begin{document}

\preprint{APS/123-QED}

\title{\textbf{Odd Capillarity}
}%

\author{Nadia Bihari Padhan}
\email{Contact author: nadia\_bihari.padhan@tu-dresden.de}
\affiliation{%
 Institute of Scientific Computing, TU Dresden, 01069 Dresden, Germany
}%
\author{Axel Voigt}%
\email{Contact author: axel.voigt@tu-dresden.de}
\affiliation{%
 Institute of Scientific Computing, TU Dresden, 01069 Dresden, Germany
}%
\affiliation{%
 Center of Systems Biology Dresden (CSBD), Pfotenhauerstr. 108, 01307 Dresden, Germany
}%
\affiliation{%
 Cluster of Excellence, Physics of Life (PoL), TU Dresden, Arnoldstr. 18, 01307 Dresden, Germany
}%

\date{\today}

\begin{abstract}
Capillarity governs interfacial relaxation and hydrodynamic motion in binary fluids through parity-even mechanical stresses. Here, we introduce \emph{odd capillarity}, a parity-breaking interfacial mechanical response, by incorporating a symmetric, nonvariational, parity-odd capillary stress into the Cahn--Hilliard--Navier--Stokes (CHNS) equations. This stress generates forces along interfaces, transverse to ordinary capillary forces, and performs mechanical work that is not balanced by changes in the composition free energy. Direct numerical simulations reveal that odd capillarity arrests domain coarsening while sustaining turbulent fluid motion without a conventional inertial energy cascade, as energy injection is predominantly balanced by viscous dissipation. In asymmetric mixtures, odd capillarity driven deformation and breakup interrupt droplet growth, producing dynamically arrested emulsions with composition hyperuniformity and suppressed large-scale vorticity fluctuations. These findings establish odd capillarity as a distinct interfacial mechanism for controlling phase separation and sustaining nonequilibrium fluid motion.

\end{abstract}

\maketitle


\textit{Introduction:} 
Capillarity converts the geometry of a fluid interface into mechanical stress~\cite{gennes2004capillarity,pomeau2006two}. This fundamental coupling governs a broad range of interfacial phenomena, from phase separation and wetting to droplet coalescence and breakup, and capillary waves ~\cite{siggia1979late,hyman2014liquid,bonn2009wetting,eggers1997nonlinear,eggers1999coalescence,lighthill2001waves}. By favoring the reduction of interfacial free energy, capillary forces drive interfacial relaxation and generate flows in the surrounding fluid, which in turn advect and deform the interfaces. This basic interplay is captured by the Cahn--Hilliard--Navier--Stokes (CHNS) equations, or Model H, in which the capillary stress couples composition gradients to fluid momentum~\cite{hohenberg1977theory,jacqmin1999calculation,abels2009diffuse,cates2018theories,padhan2025cahn}. Derived from the free energy, this stress preserves spatial parity symmetry.

The mechanical response of a material can change fundamentally when spatial reflection symmetry (parity) is broken. In chiral fluids, broken parity permits odd viscosity, a nondissipative transverse stress response~\cite{avron1998odd,banerjee2017odd}. In active solids and mechanical metamaterials, non-reciprocal interactions can give rise to odd elasticity, where elastic stresses are not derivable from a scalar energy and can perform work over cyclic deformations~\cite{scheibner2020odd,braverman2021topological}. These developments~\cite{fruchart2023odd} have primarily concerned bulk mechanical responses, whereas parity-odd capillary stresses at fluid interfaces remain comparatively unexplored. Recent theories of chiral phase separation have demonstrated persistent interfacial currents and their influence on domain morphology~\cite{langford2025phase,burekovic2026bulk,wang2026edge,blom2026mean}. These theories, however, describe dry systems, where chirality drives tangential currents of particles or composition along interfaces without momentum-conserving hydrodynamics. This raises a fundamental question: If capillarity itself acquires a parity-odd component, how does it reshape interfacial dynamics through its coupling to fluid motion?


In this Letter, we introduce \emph{odd capillarity}, a parity-breaking interfacial mechanical response, and develop a minimal hydrodynamic model incorporating a non-variational, parity-odd capillary stress within the CHNS equations. This stress generates a force transverse to the even capillary force without introducing an intrinsic torque density. Unlike the conventional capillary coupling, the odd stress can perform mechanical work that is not balanced by changes in the binary-fluid free energy. By direct numerical simulations (DNSs) we explore the possibility of odd capillarity to arrest domain coarsening and study its impact on turbulent fluid motion. We perform a scale-by-scale kinetic-energy analysis to understand the impact of odd capillarity and analyse the vorticity field with respect to hyperuniformity. Our findings establish odd capillarity as a distinct mechanism through which fluid interfaces can organize and sustain nonequilibrium motion.

\textit{Model:} We consider an incompressible binary fluid in two dimensions, described by a conserved composition field $\phi(\mathbf{x},t)$, a velocity field $\mathbf{u}(\mathbf{x},t)$ and a pressure field $p(\mathbf{x},t)$. The composition distinguishes the two coexisting phases, while its spatial gradients identify the diffuse interfaces separating them. We formulate a minimal hydrodynamic model by extending the CHNS equations~\cite{hohenberg1977theory,jacqmin1999calculation,abels2009diffuse,cates2018theories,padhan2025cahn}, retaining the usual conservation laws and introducing parity breaking solely through an additional interfacial mechanical stress. The governing equations read
\begin{align}
\partial_t\phi+\mathbf{u}\cdot\nabla\phi
&=M\nabla^2\mu,\label{eq:CH}\\
\partial_t\mathbf{u}+\mathbf{u}\cdot\nabla\mathbf{u}
&=-\nabla p+\nu\nabla^2\mathbf{u}
-\phi\nabla\mu+\nabla\cdot\boldsymbol{\Sigma}^{\rm odd},\label{eq:NS}
\end{align}
with $\nabla\cdot\mathbf{u}=0$. Here $M$ is the mobility, and $\nu$ the kinematic viscosity. We assume equal densities and viscosities in both phases, with the density set to unity. The chemical potential $\mu=\delta\mathcal F/\delta\phi$ follows from the standard Ginzburg--Landau free energy
\begin{equation}
\mathcal F[\phi]=\int_\Omega
\left[
\frac{3\sigma}{16\varepsilon}(\phi^2-1)^2+
\frac{3\sigma\varepsilon}{4}|\nabla\phi|^2
\right]d\mathbf{x},
\label{eq:free_energy}
\end{equation}
where $\sigma$ is the equilibrium surface tension and $\varepsilon$ sets the diffuse-interface thickness. Thus, the phase-separation thermodynamics remains unchanged and the additional nonequilibrium ingredient $\nabla\cdot\boldsymbol{\Sigma}^{\rm odd}$ enters only through the stress.

In ordinary CHNS equations, the capillary force $-\phi\nabla\mu$ derives from the same free energy and can be expressed, up to an isotropic pressure contribution, as the divergence of the Korteweg stress,
\begin{equation}
\Sigma_{ij}^{\rm even}
=-\frac32\sigma\varepsilon
\left[
\partial_i\phi\partial_j\phi
-\frac12|\nabla\phi|^2\delta_{ij}
\right].
\label{eq:even_stress}
\end{equation}
This stress is symmetric and parity even. Its mechanical work is compensated by changes in the binary-fluid free energy, so that capillary forces redistribute energy between interfacial and hydrodynamic degrees of freedom.

To construct an odd interfacial response, we exploit the geometry intrinsic to a diffuse interface. The composition gradient $\nabla\phi$ defines its local normal direction, while a rotation by $90^\circ$ gives the corresponding tangent. These two directions allow an interfacial stress to distinguish longitudinal and transverse mechanical responses without introducing an independent orientational field. We therefore seek a stress that is quadratic in composition gradients, symmetric under interchange of tensor indices, and odd under spatial reflection. A minimal choice satisfying these requirements is
\begin{equation}
\Sigma_{ij}^{\rm odd}
=\frac32\widetilde\sigma\varepsilon
\left[
\epsilon_{ik}\partial_k\phi\,\partial_j\phi+
\epsilon_{jk}\partial_k\phi\,\partial_i\phi
\right],
\label{eq:odd_stress}
\end{equation}
where $\epsilon_{ij}$ is the two-dimensional Levi--Civita tensor, with $\epsilon_{12}=1$. The coefficient $\widetilde\sigma$ determines the strength and handedness of odd capillarity and has the dimensions of surface tension. For $\widetilde{\sigma}=0$, the model reduces to the classical CHNS equations.

Several distinguishing mechanical properties follow from this construction. The stress is symmetric and traceless, and therefore introduces no intrinsic antisymmetric torque density. It vanishes in homogeneous phases and is concentrated in the interfacial regions where $\nabla\phi\neq0$. Most importantly, its divergence generates a force perpendicular to the passive capillary force:
\begin{equation}
\mathbf f_{\rm odd}
=2\chi\,\hat{\mathbf e}_z\times\mathbf f_{\rm even},
\qquad
\chi=\widetilde\sigma/\sigma,
\label{eq:force}
\end{equation}
where $\mathbf f_{\rm even}=\nabla\cdot\boldsymbol{\Sigma}^{\rm even}$ and $\mathbf f_{\rm odd}=\nabla\cdot\boldsymbol\Sigma^{\rm odd}$. The dimensionless chirality parameter $\chi$ measures the relative strength of odd capillarity, with $\chi=0$ recovering the achiral limit and its sign determining the handedness of the transverse response. Thus, the same composition gradients that generate ordinary capillary forcing also support an independent, parity-breaking mechanical response. For a planar interface parallel to the $x$-axis, with $\phi=\phi(y)$, the odd stress produces a tangential force density $f_x^{\rm odd}=\partial_y\Sigma_{xy}^{\rm odd}$, forming a localized force-dipole layer characterized by
$
\int_{-\infty}^{\infty}f_x^{\rm odd}\,dy=0,
\quad
\int_{-\infty}^{\infty}y f_x^{\rm odd}\,dy=-\widetilde{\sigma},
$
where the second equality follows from the equilibrium diffuse-interface profile [see Appendix~\ref{app:dipole} for the derivations]. The resulting force-dipole layer drives tangential shear across the interface, while the symmetry of the underlying stress ensures the absence of intrinsic torque density.

The proposed coupling is distinct from other nonequilibrium mechanical responses. In Active Model H~\cite{Tiribocchi2015}, an independent active stress modifies the parity-even interfacial coupling and can oppose ordinary surface tension. Its chiral extension has been developed to describe layered chiral active matter, where it yields odd elastic responses~\cite{kole2021active}. In this Letter, we instead consider a parity-odd capillary stress acting at the interfaces of a phase-separating binary fluid. This also makes it distinct from odd viscosity, which is a parity-breaking constitutive response to velocity gradients~\cite{banerjee2017odd}. 

The energetic consequences of the additional stress distinguish odd capillarity from passive interfacial relaxation. We define the mean kinetic energy $e=\langle|\mathbf u|^2\rangle_{\mathbf x}/2$, the mean free energy $f=\mathcal F/|\Omega|$, and the total mean energy $\mathcal E=e+f$, where $\langle\cdot\rangle_{\mathbf x}$ denotes a spatial average. For periodic boundary conditions, the corresponding energy balances are
\begin{subequations}\label{eq:energy_balances}
\begin{align}
\frac{de}{dt}
&=P_{\rm even}+P_{\rm odd}-D_\nu,
\label{eq:kinetic_energy}\\
\frac{df}{dt}
&=-P_{\rm even}-D_\mu,
\label{eq:free_energy_budget}\\
\frac{d\mathcal E}{dt}
&=P_{\rm odd}- D_\nu- D_\mu,
\label{eq:energy_law}
\end{align}
\end{subequations}
where $P_{\rm even}=\langle\mathbf u\cdot\mathbf f_{\rm even}\rangle_{\mathbf x}$ and $P_{\rm odd}=\langle\mathbf u\cdot\mathbf f_{\rm odd}\rangle_{\mathbf x}$ are the powers transferred to the fluid by even and odd capillary forces, respectively. The viscous and diffusive dissipation rates are $ D_\nu=-\nu\langle\mathbf u\cdot\nabla^2\mathbf u\rangle_{\mathbf x}$ and $ D_\mu=M\langle|\nabla\mu|^2\rangle_{\mathbf x}$, respectively, with $ D_\nu, D_\mu\geq0$ [see Appendix~\ref{app:energy-balance} for detailed derivations]. Unlike the passive capillary contribution, whose mechanical work is balanced by changes in the composition free energy, the odd stress produces an independent power term $P_{\rm odd}$, which can have either sign. The model therefore permits sustained mechanical energy injection through stresses localized at fluid interfaces, even though the composition dynamics retains its conventional Cahn--Hilliard form.
%
\begin{figure*}[t]
    \centering
    \includegraphics[width=\textwidth]{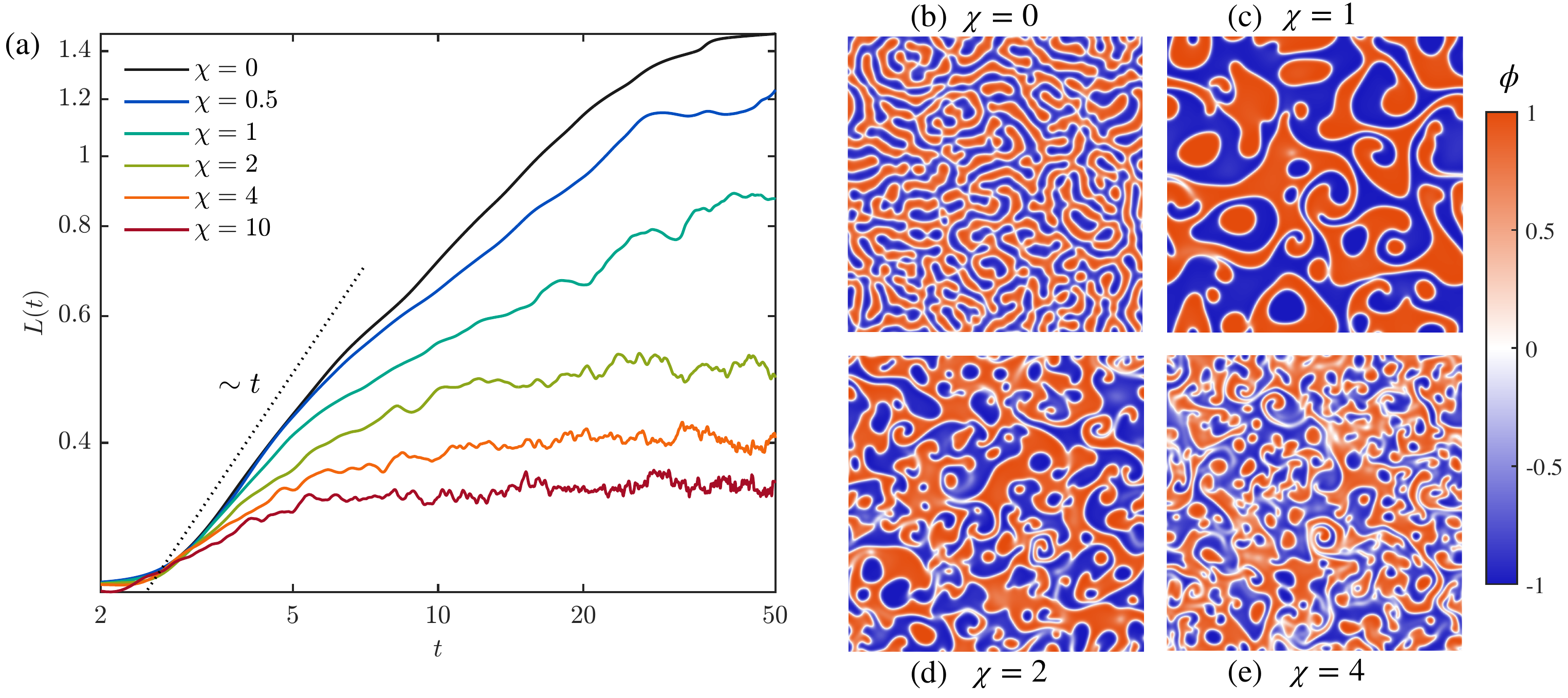}

\caption{
\textbf{Odd capillarity slows and arrests domain coarsening.}
(a) Evolution of the characteristic coarsening length $L(t)$ for different chirality parameter $\chi=\widetilde{\sigma}/\sigma$, with $\sigma=1$. For $\chi=0$ (black solid line), the ordinary CHNS model exhibits approximately linear hydrodynamic coarsening, $L(t)\sim t$ (black dotted line). Increasing $\chi$ progressively suppresses domain growth, with coarsening arrest observed for $\chi=2$, $4$, and $10$. The arrested length decreases with increasing $\chi$.
(b)--(e) Representative pseudocolor plots of the composition field $\phi$: (b) $\chi=0$ at $t=4$; (c)--(e) $\chi=1$, $2$, and $4$, respectively, at $t=20$, illustrating the increasingly fine domain structures.}
\label{fig:coarsening}
\end{figure*}

\textit{Phase ordering and coarsening arrest:}
We first investigate how odd capillarity modifies the phase-ordering dynamics of a binary fluid. We perform DNSs of Eqs. \eqref{eq:CH} and \eqref{eq:NS} using a Fourier pseudospectral method~\cite{padhan2025cahn} in a two-dimensional periodic square of side $2\pi$, with $1024^2$ collocation points [see Appendix~\ref{app:DNS} for numerical details]. Starting from a nearly homogeneous mixture, $\phi(\mathbf{x},0)=\phi_0+\delta\phi(\mathbf{x})$, where $\phi_0$ is the mean composition and $\delta\phi$ is a small-amplitude random perturbation, we examine the evolution of the characteristic domain size. We first consider a symmetric mixture, $\phi_0=0$, and define the coarsening length scale as
\begin{equation}
L(t)=2\pi\frac{\sum_k S(k,t)}{\sum_k kS(k,t)},
\label{eq:length_scale}
\end{equation}
where $S(k,t)=\sum_{k\leq|\mathbf{k}'|<k+\Delta k}|\widehat{\delta\phi}(\mathbf{k}',t)|^2$ is the shell-integrated composition spectrum, with $\delta\phi=\phi-\phi_0$. Here, $\widehat{\phi}(\mathbf{k}',t)$ denotes the spatial Fourier coefficient of $\phi$, $\mathbf{k}'$ is the wavevector, $k$ is the wavenumber defining the shell, and $\Delta k$ is its width.
%
\begin{figure*}[t]
    \centering
    \includegraphics[width=\textwidth]{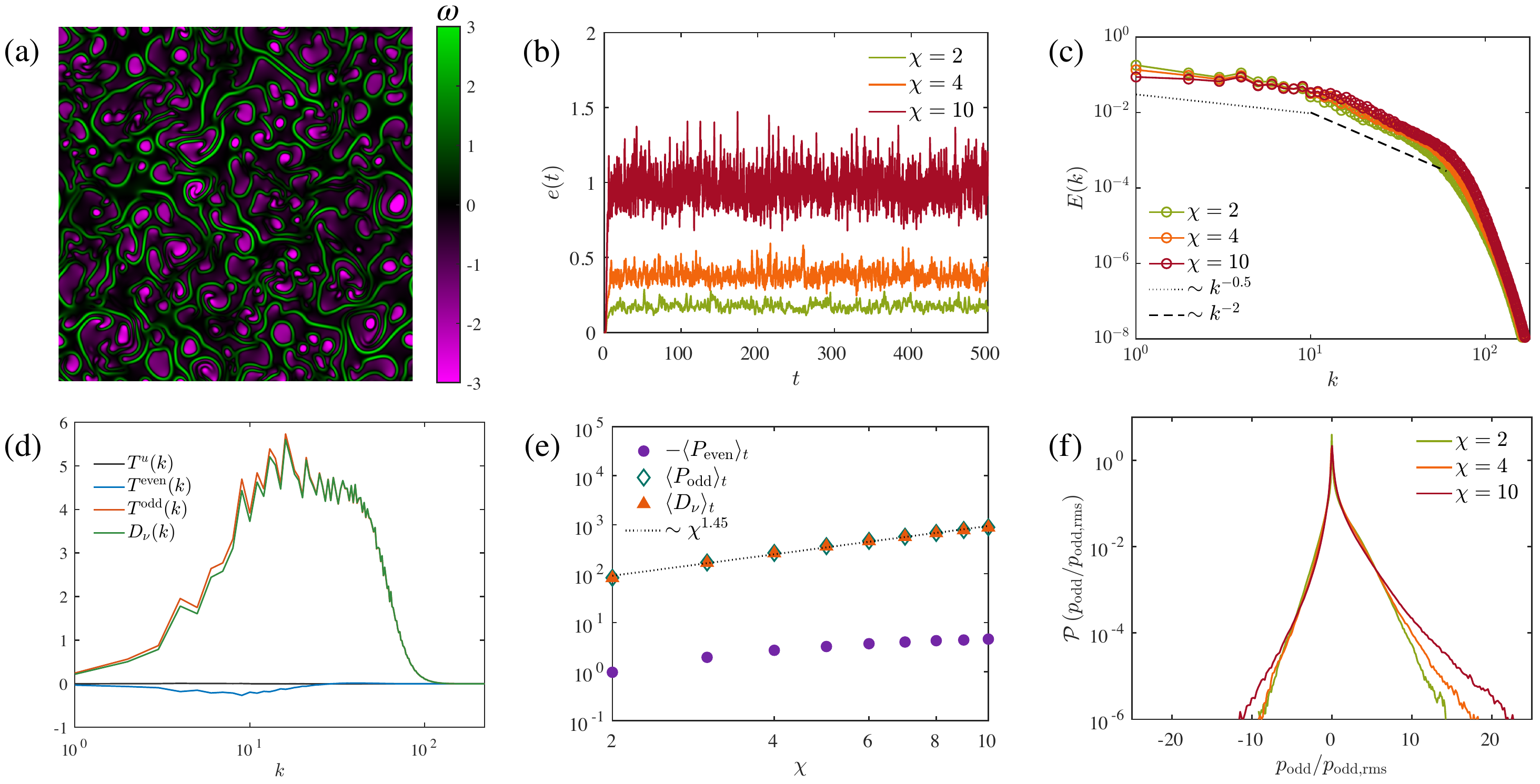}
    \caption{
\textbf{Odd-capillary-driven turbulence.}
(a) Representative pseudocolor plot of the scalar vorticity field $\omega=(\nabla\times\mathbf u)\cdot\hat{\mathbf e}_z$, normalized by its root-mean-square value $\omega_{\rm rms}$, for $\chi=4$ at $t=400$.
(b) Time evolution of the spatially averaged kinetic energy $e(t)$ for $\chi=2$, $4$, and $10$.
(c) Time-averaged kinetic-energy spectra $E(k)$, normalized by $\langle e(t)\rangle_t$, for the same values of $\chi$. The dotted and dashed lines indicate the reference scalings $k^{-1/2}$ and $k^{-2}$, respectively. The spectra are averaged over 400 snapshots in the statistically stationary state.
(d) Spectral contributions $T^u(k)$, $T^{\rm even}(k)$, $T^{\rm odd}(k)$, and $D_\nu(k)$ to the kinetic-energy balance [Eq.~\eqref{eq:scale_by_scale}] for $\chi=4$, plotted on a logarithmic wavenumber axis. The inertial transfer is negligible, while odd-capillary energy injection is predominantly balanced by viscous dissipation; the even capillary contribution is approximately two orders of magnitude smaller.
(e) Mean odd-capillary power $\langle P_{\rm odd}\rangle_t$, viscous dissipation $\langle\mathcal D_\nu\rangle_t$, and even capillary power $-\langle P_{\rm even}\rangle_t$ versus $\chi$. The dotted line indicates the scaling $\langle P_{\rm odd}\rangle_t\sim\chi^{1.45}$. The ratio $|\langle P_{\rm even}\rangle_t|/\langle P_{\rm odd}\rangle_t\sim10^{-2}$.
(f) Probability density functions of the local odd-capillary power $p_{\rm odd}(\mathbf x,t)=\mathbf u\cdot\mathbf f_{\rm odd}$, normalized by its root-mean-square value, for $\chi=2$, $4$, and $10$.
}
\label{fig:turbulence}
\end{figure*}
Figure~\ref{fig:coarsening}(a) shows the temporal evolution of $L(t)$ for different values of chirality parameter $\chi$, while Figs. \ref{fig:coarsening}(b)–(e) display representative pseudocolor plots of $\phi$. In the achiral limit ($\chi=0$), the mixture undergoes spinodal decomposition, forming interconnected domains that grow through diffusive transport and hydrodynamic motion [Fig. \ref{fig:coarsening}(b)]. In the observed scaling regime, $L(t)\sim t$, consistent with the classical viscous hydrodynamic growth law of binary-fluid mixtures~\cite{siggia1979late,padhan2025suppression}. Introducing odd capillarity significantly alters these dynamics. For $\chi=0.5$ and $1$, coarsening continues but becomes progressively slower, with the growth rate decreasing as $\chi$ increases. The corresponding interfaces exhibit pronounced deformations and irregular domain morphologies [Fig. \ref{fig:coarsening}(c)]. Upon increasing $\chi$ to $2$, the characteristic length ceases to grow systematically and fluctuates around a statistically stationary value, indicating coarsening arrest. At stronger odd capillarity ($\chi=4, 10$), the arrested state persists with a further reduction in the characteristic domain size [Figs. \ref{fig:coarsening}(d) and \ref{fig:coarsening}(e)]. 
In the arrested state, the interfaces keep evolving through
diffusive growth, reconnection, neck thinning, and breakup,
and spiral-like structures form. The state is dynamically sustained rather than frozen. The free-energy balance [Eq.~\ref{eq:free_energy_budget}] provides an energetic interpretation of this behaviour. In a non-equilibrium steady state,
$
\langle P_{\rm even}\rangle_t
=-\left\langle
{D_{\mu}}
\right\rangle_t<0.
$
Thus, on average, the flow must continuously do work against capillarity
and regenerate interfacial free energy at the rate it is
dissipated by diffusion. In contrast, for the classical CHNS equations in the Stokes limit, $P_{\rm even}=D_\nu\geq0$, and the free energy decreases monotonically. Odd capillarity therefore sustains a dynamically arrested state in which interfaces continually evolve despite the absence of systematic domain growth. Coarsening arrest in hydrodynamically coupled binary fluids has previously been observed under externally driven turbulent flows~\cite{berti2005arrest,perlekar2014spinodal,perlekar2017two}, through contractile interfacial stresses in Active Model H~\cite{active_model_H,padhan2024novel,padhan2025suppression}, and through bulk active flows~\cite{marchetti2022suppressed,backofen2026self}. Arrested phase separation has also been reported in active-spinner fluids and non-reciprocal binary-fluid turbulence~\cite{maji2025emergent,maji2026non}. The described mechanism differs from these examples as odd capillarity suppresses coarsening through self-generated tangential interfacial flows and the resulting hydrodynamic deformation of domains.

%
\begin{figure*}[t]
    \centering
    \includegraphics[width=\textwidth]{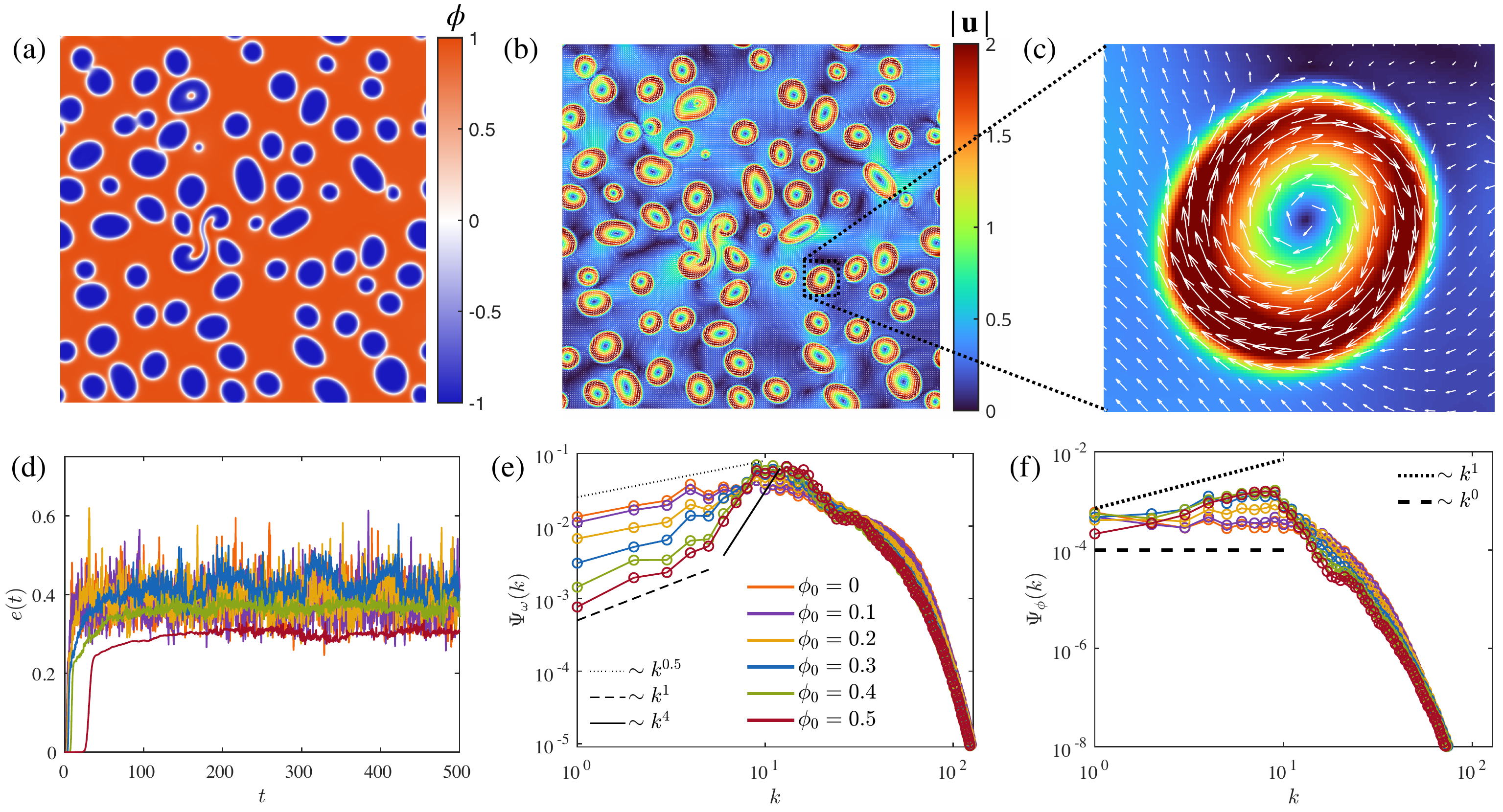}

    \caption{
\textbf{Arrested droplet growth and hyperuniformity.}
All results are for $\chi=4$.
(a) Representative pseudocolor plot of the composition field $\phi$ for $\phi_0=0.4$ at $t=300$.
(b) Corresponding pseudocolor plot of the velocity magnitude $|\mathbf u|$, overlaid with velocity vectors.
(c) Magnified view of an individual droplet in (b), highlighting the interfacial flow and velocity vectors.
(d) Time evolution of the spatially averaged kinetic energy $e(t)$ for $\phi_0=0$, $0.1$, $0.2$, $0.3$, $0.4$, and $0.5$.
(e) Time-averaged vorticity spectral density $\Psi_\omega(k)$ for different $\phi_0$, showing suppressed long-wavelength fluctuations and the emergence of hyperuniformity. The reference slopes indicate the scalings $k^{1/2}$, $k$, and $k^4$.
(f) Time-averaged composition spectral density $\Psi_\phi(k)$ for the same values of $\phi_0$. The reference slopes $k^0$ and $k$ illustrate the transition from nonhyperuniform to hyperuniform composition fluctuations. The color legend in (e) applies to (d)–(f).
}
\label{fig:droplets}
\end{figure*}

\textit{Odd-capillary-driven turbulence:} The persistent interfacial deformations and coarsening arrest raise a fundamental question about the origin of the accompanying fluid motion. Can odd capillarity itself sustain complex flows in the absence of appreciable inertia or external forcing? To address this question, we examine the kinetic-energy spectrum and the mechanisms of energy transfer across scales. The shell-integrated kinetic-energy spectrum, $E(k,t)=\sum_{k\leq|\mathbf{k}'|<k+\Delta k}|\widehat{\mathbf u}(\mathbf{k}',t)|^2$, satisfies the scale-by-scale energy balance~\cite{perlekar2019kinetic,alexakis2018cascades,verma2019energy}
\begin{equation}
\partial_t E(k,t)
=T^u(k,t)+T^{\rm even}(k,t)
+T^{\rm odd}(k,t)-D_\nu(k,t),
\label{eq:scale_by_scale}
\end{equation}
where $T^u$, $T^{\rm even}$, and $T^{\rm odd}$ denote the contributions from nonlinear advection (inertia), even capillary forcing, and odd-capillary forcing, respectively, and $D_\nu$ is the viscous dissipation spectrum [see Appendix~\ref{app:spectral-energy} for the definitions]. Hereafter, $E(k)=\langle E(k,t)\rangle_t$, $T^\alpha(k)=\langle T^\alpha(k,t)\rangle_t$, and $D_\nu(k)=\langle D_\nu(k,t)\rangle_t$ denote their statistically stationary time averages, with $\alpha=u,{\rm even},{\rm odd}$. In the dynamically arrested regime, the kinetic energy fluctuates around a statistically stationary mean [Fig.~\ref{fig:turbulence}(b)], while its spectrum extends over a broad range of wavenumbers [Fig.~\ref{fig:turbulence}(c)], with $E(k) \sim k^{-0.5}$ for $k < 2\pi/L_c$ and $E(k) \sim k^{-2}$ for $k > 2\pi/L_c$, where $L_c$ is the characteristic arrested domain size. The scale-by-scale energy balance reveals that $T^{\rm odd}(k)$ is predominantly positive, demonstrating that odd capillarity injects mechanical energy directly into the fluid across a range of scales [Fig.~\ref{fig:turbulence}(d)]. The injected energy is balanced primarily by viscous dissipation $D_{\nu}(k)$, while the nonlinear advective transfer $T^{u}$ is negligible. The even capillary contribution is comparatively weak and changes sign with wavenumber, while the nonlinear advective transfer $T^u(k)$ is negligible. In the non-equilibrium stationary state, $\langle\partial_t E(k,t)\rangle_t=0$, so Eq.~\ref{eq:scale_by_scale} gives $T^{\rm odd}(k)+T^{\rm even}(k)\simeq D_\nu(k)$. Since the even capillary contribution is comparatively small, $T^{\rm odd}(k)\simeq D_\nu(k)$, indicating that energy injected by odd capillarity is dissipated predominantly at the same scales by viscosity, without appreciable transfer through an inertial cascade.

Since the advective transfer only redistributes energy among scales, $\sum_kT^u(k)=0$ for incompressible periodic flow, summing Eq.~\ref{eq:scale_by_scale} over all wavenumbers recovers the global kinetic-energy balance Eq.~\ref{eq:kinetic_energy}, with $e(t)=\sum_k E(k,t)$, $P_{\rm odd}(t)=\sum_k T^{\rm odd}(k,t)$, $P_{\rm even}(t)=\sum_k T^{\rm even}(k,t)$, and $ D_\nu(t)=\sum_k D_\nu(k,t)$. Figure~\ref{fig:turbulence}(e) shows the time-averaged terms in the non-equilibrium steady state. We find $\langle P_{\rm odd}\rangle_t>0$ and $\langle P_{\rm even}\rangle_t < 0$, which indicates that odd capillarity injects kinetic energy globally, while the even capillary force converts a small fraction of it into composition free energy, $-\langle P_{\rm even}\rangle_t/\langle P_{\rm odd}\rangle_t \approx 10^{-2}$. This fraction balances diffusive dissipation, $\langle P_{\rm even}\rangle_t =-\left\langle {D_{\mu}} \right\rangle_t<0$, as required by the free-energy balance [Eq.~\ref{eq:free_energy_budget}]. The remainder is dissipated by viscosity, so that $\langle P_{\rm odd}\rangle_t=\langle D_\nu\rangle_t
+\langle D_\mu\rangle_t\simeq\langle D_\nu\rangle_t$.
The mean odd capillary power scales as $\langle P_{\rm odd}\rangle_t\sim\chi^{1.45}$. Although odd capillarity injects energy on average, the local power density $p_{\rm odd}(\mathbf x,t)=\mathbf u\cdot\mathbf f_{\rm odd}$ fluctuates between positive and negative values. Its probability density $\mathcal P(p_{\rm odd}/p_{{\rm odd,rms}})$ exhibits tails of both signs, with a stronger positive tail [Fig.~\ref{fig:turbulence}(f)]. The distributions broaden with increasing $\chi$, indicating increasingly pronounced relative fluctuations in local energy transfer. Thus, odd capillarity acts as a net energy source while locally both injecting and extracting kinetic energy. In contrast to externally sustained interface-induced turbulence in viscous binary fluids~\cite{padhan2024interface}, the present turbulent dynamics are powered directly by parity-odd stresses at evolving fluid interfaces.

\textit{Arrested droplet growth and hyperuniformity:} Having established coarsening arrest and sustained fluid motion in symmetric mixtures, we now examine phase separation in asymmetric mixtures with $\phi_0=0.1,0.2,0.3,0.4$, and $0.5$. Figure~\ref{fig:droplets}(a) shows a representative composition field for $\phi_0=0.4$ and $\chi=4$, revealing a dynamically evolving droplet-rich state. Unlike passive binary-fluid mixtures, where droplets grow through Ostwald ripening and coalescence, odd capillarity interrupts their continued growth. Although smaller droplets shrink and disappear while larger ones grow through Ostwald ripening, the latter undergo interfacial instabilities that lead to deformation and breakup. The associated velocity fields [Figs.~\ref{fig:droplets}(b) and (c)] reveal strong tangential flows near the droplet interfaces, which also hinder coalescence between neighboring droplets. Thus, the competition between diffusive growth, coalescence, and odd capillary driven breakup maintains a dynamically arrested droplet-rich state.

Figure~\ref{fig:droplets}(d) shows that the mean kinetic energy $e(t)$ approaches a statistically stationary state for different $\phi_0$. Its temporal fluctuations become progressively weaker with increasing $\phi_0$, accompanied by increasingly localized flows associated with individual droplets. To characterize the resulting large-scale vorticity fluctuations, we examine its hyperuniformity, which is characterized by anomalously suppressed large scale fluctuations~\cite{torquato2018hyperuniform}. We consider the time-averaged spectral density of the vorticity field, $\omega=(\nabla\times\mathbf u)\cdot\hat{\mathbf e}_z$, defined as
$
\Psi_\omega(k)=\langle\frac{1}{N_k}
\sum_{k\leq|\mathbf k'|<k+\Delta k}
|\widehat\omega(\mathbf k',t)|^2\rangle_t,
$
where $N_k$ is the number of Fourier modes in the shell $k\leq|\mathbf k'|<k+\Delta k$. In two dimensions, $\Psi_\omega(k)\propto kE(k)$~\cite{backofen2024nonequilibrium}, which relates the vorticity spectral density to the kinetic-energy spectrum. For symmetric mixtures ($\phi_0=0$), we find $\Psi_\omega(k)\sim k^{0.5}$ at small wavenumbers [Fig.~\ref{fig:droplets}(e)], consistent with $E(k)\sim k^{-0.5}$ discussed earlier [Fig.~\ref{fig:turbulence}(c)]. With increasing $\phi_0$, this scaling gives way to two distinct regimes, $\Psi_\omega(k)\sim k$ at smaller wavenumbers and $\Psi_\omega(k)\sim k^4$ at relatively higher wavenumbers. These positive exponents indicate hyperuniform vorticity fluctuations, with increasingly strong suppression at large scales. 

The composition field also exhibits a marked change in its large-scale fluctuations. Figure~\ref{fig:droplets}(f) shows the time-averaged spectral density $\Psi_\phi(k)$, defined analogously to $\Psi_\omega(k)$. For symmetric mixtures, $\Psi_\phi(k)\sim k^0$ at small wavenumbers, indicating nonhyperuniform composition fluctuations. In contrast, the droplet-rich states exhibit $\Psi_\phi(k)\sim k$, revealing the emergence of composition hyperuniformity. As vortical flows become increasingly localized around individual droplets, their spatial arrangement influences the large-scale distribution of vorticity, providing a physical connection between the hyperuniformity of the composition and vorticity fields. Hydrodynamically interrupted droplet growth has also been reported in Active Model H with contractile interfacial stresses~\cite{singh2019droplet}. However, such activity can suppress composition hyperuniformity and produce anti-hyperuniform vorticity fluctuations, $\Psi_\omega(k)\sim k^{-2/3}$~\cite{padhan2025suppression}. In contrast, odd capillarity arrests droplet growth while promoting composition hyperuniformity and suppressing large-scale vorticity fluctuations.

\textit{Conclusions and outlook:} We have shown that capillarity admits a parity-odd counterpart, \emph{odd capillarity}, through which fluid interfaces generate transverse forces and sustain non-equilibrium motion. We introduced this response within the CHNS equations by supplementing the ordinary Korteweg stress with a symmetric, parity-odd capillary stress. The resulting localized force-dipole structure drives interfacial shear flows that, in turn, deform and reorganize the interfaces themselves. This feedback fundamentally changes phase separation, progressively slowing and ultimately arresting coarsening.

We have identified a sustained energy-transfer cycle underlying coarsening arrest. Odd-capillary stresses inject kinetic energy into the fluid, most of which is dissipated by viscosity, while a smaller fraction is transferred through ordinary capillary stresses to replenish the composition free energy lost through diffusion. This energy transfer sustains interfacial rearrangements and prevents continued domain growth. We have further demonstrated that the arrested state exhibits odd-capillary-driven turbulence, characterized by persistent kinetic-energy fluctuations and broadband kinetic energy spectra. Our scale-by-scale analysis reveals that energy injection is predominantly balanced by viscous dissipation at the same scales, with negligible inertial transfer. In asymmetric mixtures, the same interfacial feedback interrupts droplet growth: although Ostwald ripening persists, growing droplets deform and break under odd-capillary forcing, while tangential interfacial flows hinder coalescence. The resulting dynamically arrested droplet-rich states exhibit composition hyperuniformity and strongly suppressed large-scale vorticity fluctuations.

Odd capillarity opens new possibilities for revisiting classical capillary phenomena in systems with broken parity. Unlike chiral surface waves arising from bulk odd viscosity~\cite{abanov2018odd} or odd slip at chiral active boundaries~\cite{hosaka2026odd}, parity breaking here enters directly through the capillary stress. A natural next step is to develop a systematic sharp-interface theory to establish the mechanisms of interfacial deformation and instability. Such a theory would provide a foundation for exploring whether odd capillarity can destabilize interfaces even in the absence of inertia, how it modifies capillary waves and Rayleigh--Plateau breakup~\cite{barker2023fluctuating}, and how it interacts with external forces. In particular, the motion of rising or falling droplets under gravity presents an intriguing problem, complementing studies of deformable droplets with bulk odd viscosity~\cite{franca2026odd}. Deriving odd capillarity from microscopic models of chiral active constituents remains an open challenge. Coarse-graining approaches that have yielded interfacial tension and odd surface flows~\cite{langford2025phase} and odd transport coefficients~\cite{maire2026kinetic} for chiral active particles offer natural routes. Experiments provide different routes to generating chiral flows near fluid interfaces. In a colloidal fluid driven by external torques, the particles develop persistent edge flows, while the free surface exhibits waves and instabilities~\cite{soni2019odd}. Another possibility involves motile constituents interacting directly with fluid interfaces. Bacteria adsorbed at oil--water interfaces can drive droplet rotation, with angular speed decreasing with droplet size~\cite{dewangan2019rotating}. Bacteria trapped at fluid interfaces also exhibit circular swimming trajectories~\cite{deng2020motile} and generate localized hydrodynamic force dipoles~\cite{deng2023interfacial}. These observations motivate investigating whether microscopic chirality can produce the parity-odd interfacial stress proposed here, providing a possible experimental route to odd capillarity. These observations suggest possible experimental routes to odd capillarity, although whether microscopic chirality generates the proposed parity-odd interfacial stress remains to be established. Incorporating substrate friction, which is relevant to many experimental systems, would extend our model to settings with momentum loss. More broadly, from ordinary surface tension in passive fluids to negative mechanical tension in active fluids~\cite{singh2019droplet}, capillarity has acquired new roles in interfacial dynamics. Odd capillarity introduces a distinct possibility: fluid interfaces that retain their ordinary surface tension while generating handed, tangential flows that inject mechanical energy and organize the surrounding fluid.

\textit{Acknowledgments:} We acknowledges the support by the German Research Foundation (DFG) through the project ``Analysing structure–property relations
in equilibrium and non-equilibrium hyperuniform systems"
(project number VO 899/32-1). We further acknowledge
computing resources at JSC under grant "MORPH" and at ZIH
under grant "WIR".


\appendix

\section{Tangential force-dipole structure of odd capillarity}
\label{app:dipole}
To characterize the interfacial forcing generated by the odd
stress, consider a planar interface parallel to the $x$-axis,
with $\phi=\phi_0(y)$. For $\epsilon_{xy}=1$, the odd stress
reduces to
\begin{equation}
\boldsymbol{\Sigma}^{\rm odd}
=\frac{3}{2}\widetilde{\sigma}\varepsilon
[\phi_0'(y)]^2
\begin{pmatrix}
0 & 1\\
1 & 0
\end{pmatrix}.
\end{equation}
Its divergence yields a purely tangential force density,
\begin{equation}
f_x^{\rm odd}
=\frac{3}{2}\widetilde{\sigma}\varepsilon
\frac{d}{dy}[\phi_0'(y)]^2,
\qquad
f_y^{\rm odd}=0.
\end{equation}
For the equilibrium diffuse-interface profile
$\phi_0(y)=\tanh[y/(2\varepsilon)]$, this becomes
\begin{equation}
f_x^{\rm odd}(y)
=-\frac{3\widetilde{\sigma}}{4\varepsilon^2}
\operatorname{sech}^4
\left(\frac{y}{2\varepsilon}\right)
\tanh\left(\frac{y}{2\varepsilon}\right).
\end{equation}
The force density is antisymmetric with respect to the interface.
Its zeroth moment vanishes, whereas its first normal moment
remains finite:
\begin{equation}
\int_{-\infty}^{\infty}f_x^{\rm odd}\,dy=0,
\qquad
\int_{-\infty}^{\infty}y f_x^{\rm odd}\,dy
=-\widetilde{\sigma}.
\end{equation}
The latter follows by integration by parts and the identity
$\int_{-\infty}^{\infty}[\phi_0'(y)]^2\,dy
=2/(3\varepsilon)$.

Thus, the odd stress produces a localized tangential
force-dipole layer with zero resultant force and a nonzero
first moment.

\section{Derivation of the energy balances}
\label{app:energy-balance}

We derive the kinetic-, free-, and total-energy balances [Eqs.~\eqref{eq:energy_balances}] for incompressible flow with periodic boundary conditions. The derivation also applies to boundary conditions for which the corresponding boundary terms vanish. We define the spatial average as
\begin{equation}
\langle g\rangle_{\mathbf x}
=\frac{1}{|\Omega|}\int_\Omega g(\mathbf x)\,d\mathbf x.
\end{equation}
The mean kinetic energy, mean free energy, and total mean energy are
\begin{equation}
e=\frac12\langle|\mathbf u|^2\rangle_{\mathbf x},
\qquad
f=\frac{\mathcal F[\phi]}{|\Omega|},
\qquad
\mathcal E=e+f,
\end{equation}
where the density is constant and set to $\rho=1$. The chemical potential is defined by $\mu=\delta\mathcal F/\delta\phi$.

\textit{Free-energy balance.} Since $\mu=\delta\mathcal F/\delta\phi$, the time derivative of the mean free energy is
\begin{equation}
\frac{df}{dt}
=\langle\mu\,\partial_t\phi\rangle_{\mathbf x}.
\end{equation}
Substituting the Cahn--Hilliard equation,
\begin{equation}
\partial_t\phi=-\mathbf u\cdot\nabla\phi+M\nabla^2\mu,
\end{equation}
gives
\begin{align}
\frac{df}{dt}
&=-\langle\mu\,\mathbf u\cdot\nabla\phi\rangle_{\mathbf x}
+M\langle\mu\nabla^2\mu\rangle_{\mathbf x}\nonumber\\
&=\langle\phi\,\mathbf u\cdot\nabla\mu\rangle_{\mathbf x}
-M\langle|\nabla\mu|^2\rangle_{\mathbf x},
\label{eq:app-free-energy-rate}
\end{align}
where we have used incompressibility, periodicity, and integration by parts.

The ordinary capillary force appearing in the Navier--Stokes equation is $-\phi\nabla\mu$. Its mean mechanical power is
\begin{equation}
P_{\rm even}
=-\langle\phi\,\mathbf u\cdot\nabla\mu\rangle_{\mathbf x}.
\end{equation}
Equivalently, since $-\phi\nabla\mu$ and $\mathbf f_{\rm even}=\nabla\cdot\boldsymbol\Sigma^{\rm even}$ differ only by a gradient, whose mechanical work vanishes for incompressible periodic flow,
\begin{equation}
P_{\rm even}
=\langle\mathbf u\cdot\mathbf f_{\rm even}\rangle_{\mathbf x}.
\end{equation}
Defining the nonnegative diffusive dissipation rate as
\begin{equation}
 D_\mu=M\langle|\nabla\mu|^2\rangle_{\mathbf x},
\end{equation}
we obtain
\begin{equation}
\frac{df}{dt}=-P_{\rm even}- D_\mu.
\label{eq:app-free-energy-budget}
\end{equation}

\textit{Kinetic-energy balance.} Multiplying the Navier--Stokes equation [Eq.~\eqref{eq:NS}] by $\mathbf u$ and taking the spatial average gives
\begin{align}
\frac{de}{dt}
=&-\langle\mathbf u\cdot\nabla p\rangle_{\mathbf x}
+\nu\langle\mathbf u\cdot\nabla^2\mathbf u\rangle_{\mathbf x}
\nonumber\\
&+\langle\mathbf u\cdot\mathbf f_{\rm even}\rangle_{\mathbf x}
+\langle\mathbf u\cdot\mathbf f_{\rm odd}\rangle_{\mathbf x},
\label{eq:app-kinetic-start}
\end{align}
where the nonlinear advective contribution vanishes because
\begin{align}
\langle\mathbf u\cdot(\mathbf u\cdot\nabla)\mathbf u\rangle_{\mathbf x}
&=\left\langle\mathbf u\cdot\nabla\frac{|\mathbf u|^2}{2}\right\rangle_{\mathbf x}\nonumber\\
&=\left\langle\nabla\cdot\left(\frac{|\mathbf u|^2}{2}\mathbf u\right)\right\rangle_{\mathbf x}
=0.
\end{align}
Likewise, the pressure term vanishes:
\begin{equation}
\langle\mathbf u\cdot\nabla p\rangle_{\mathbf x}
=\langle\nabla\cdot(p\mathbf u)\rangle_{\mathbf x}
-\langle p\,\nabla\cdot\mathbf u\rangle_{\mathbf x}
=0.
\end{equation}

For constant viscosity, integration by parts gives
\begin{align}
\nu\langle\mathbf u\cdot\nabla^2\mathbf u\rangle_{\mathbf x}
&=-\nu\langle|\nabla\mathbf u|^2\rangle_{\mathbf x}\nonumber\\
&=-2\nu\langle\mathbf S:\mathbf S\rangle_{\mathbf x},
\end{align}
where $\mathbf S=(\nabla\mathbf u+\nabla\mathbf u^T)/2$. We therefore define the positive viscous dissipation rate as
\begin{equation}
\mathcal D_\nu
=-\nu\langle\mathbf u\cdot\nabla^2\mathbf u\rangle_{\mathbf x}
=2\nu\langle\mathbf S:\mathbf S\rangle_{\mathbf x}
\geq0.
\end{equation}

The mechanical power supplied by the odd capillary force is
\begin{equation}
P_{\rm odd}
=\langle\mathbf u\cdot\mathbf f_{\rm odd}\rangle_{\mathbf x},
\qquad
\mathbf f_{\rm odd}=\nabla\cdot\boldsymbol\Sigma^{\rm odd}.
\end{equation}

Collecting the contributions, the kinetic-energy balance becomes
\begin{equation}
\frac{de}{dt}
=P_{\rm even}+P_{\rm odd}- D_\nu.
\label{eq:app-kinetic-budget}
\end{equation}

\textit{Total-energy balance.} Adding Eqs.~\eqref{eq:app-free-energy-budget} and \eqref{eq:app-kinetic-budget}, the ordinary capillary power cancels exactly between the kinetic- and free-energy balances. Thus,
\begin{equation}
\frac{d\mathcal E}{dt}
=P_{\rm odd}- D_\nu- D_\mu.
\label{eq:app-total-energy-budget}
\end{equation}
This recovers the total-energy balance in Eq.~\eqref{eq:energy_law}. The ordinary capillary force mediates energy exchange between the fluid motion and the composition field, whereas the odd capillary stress contributes an independent, sign-indefinite power to the total-energy budget.

For $\widetilde\sigma=0$, the odd power vanishes, and the total mean energy satisfies
\begin{equation}
\frac{d\mathcal E}{dt}
=- D_\nu- D_\mu\leq0,
\end{equation}
recovering the dissipative structure of passive Cahn--Hilliard--Navier--Stokes dynamics.

\section{Spectral kinetic-energy balance}
\label{app:spectral-energy}

We derive the scale-by-scale kinetic-energy balance used in the main text, following the spectral-transfer framework for Cahn--Hilliard--Navier--Stokes turbulence~\cite{perlekar2019kinetic,padhan2024novel,padhan2024interface}. For constant density $\rho=1$, the incompressible Navier--Stokes equation is
\begin{equation}
\partial_t\mathbf u+(\mathbf u\cdot\nabla)\mathbf u
=-\nabla p+\nu\nabla^2\mathbf u
+\mathbf f_{\rm even}+\mathbf f_{\rm odd},
\label{eq:app-spectral-ns}
\end{equation}
where $\mathbf f_{\rm even}=\nabla\cdot\boldsymbol\Sigma^{\rm even}$ and $\mathbf f_{\rm odd}=\nabla\cdot\boldsymbol\Sigma^{\rm odd}$. The ordinary capillary force is equivalent to $-\phi\nabla\mu$ up to a gradient, which does not contribute to the kinetic-energy balance of incompressible periodic flow.

For an $N\times N$ periodic grid, we define the normalized discrete Fourier coefficients as
\begin{equation}
\widehat{\mathbf u}(\mathbf k,t)
=\frac{1}{N^2}\sum_{i,j=0}^{N-1}
\mathbf u(\mathbf x_{ij},t)
e^{-i\mathbf k\cdot\mathbf x_{ij}}.
\label{eq:app-fourier}
\end{equation}
The same convention applies to all other fields. Here $\mathbf k=(2\pi/L)(n_x,n_y)$ is the physical wavevector for a square domain of side $L$. The Fourier-transformed momentum equation can be written as
\begin{align}
\partial_t\widehat{\mathbf u}(\mathbf k,t)
=&-\mathsf P(\mathbf k)\cdot
\widehat{(\mathbf u\cdot\nabla)\mathbf u}(\mathbf k,t)
-\nu|\mathbf k|^2\widehat{\mathbf u}(\mathbf k,t)
\nonumber\\
&+\mathsf P(\mathbf k)\cdot
\left[
\widehat{\mathbf f}_{\rm even}(\mathbf k,t)
+\widehat{\mathbf f}_{\rm odd}(\mathbf k,t)
\right],
\label{eq:app-fourier-ns}
\end{align}
where
\begin{equation}
\mathsf P_{ij}(\mathbf k)
=\delta_{ij}-\frac{k_i k_j}{|\mathbf k|^2}
\end{equation}
is the incompressible projection operator for $\mathbf k\neq0$. It removes the pressure contribution and ensures that $\mathbf k\cdot\widehat{\mathbf u}(\mathbf k,t)=0$.

The shell-integrated kinetic-energy spectrum is
\begin{equation}
E(k,t)=\frac12
\sum_{\mathbf k'\in\mathcal S_k}
|\widehat{\mathbf u}(\mathbf k',t)|^2,
\label{eq:app-energy-spectrum}
\end{equation}
where $\mathcal S_k=\{\mathbf k':k\leq|\mathbf k'|<k+\Delta k\}$ denotes a wavenumber shell of width $\Delta k$. Multiplying Eq.~\eqref{eq:app-fourier-ns} by $\widehat{\mathbf u}^{\,*}$, taking the real part, and summing over the shell gives
\begin{equation}
\partial_tE(k,t)
=T^u(k,t)+T^{\rm even}(k,t)
+T^{\rm odd}(k,t)-D_\nu(k,t).
\label{eq:app-spectral-budget}
\end{equation}
The nonlinear advective transfer is
\begin{align}
T^u(k,t)
=-\sum_{\mathbf k'\in\mathcal S_k}
\operatorname{Re}\Big[
\widehat{\mathbf u}^{\,*}(\mathbf k',t)\cdot
\mathsf P(\mathbf k')\cdot
\widehat{(\mathbf u\cdot\nabla)\mathbf u}(\mathbf k',t)
\Big].
\label{eq:app-nonlinear-transfer}
\end{align}
The ordinary and odd capillary contributions are
\begin{align}
T^{\rm even}(k,t)
&=\sum_{\mathbf k'\in\mathcal S_k}
\operatorname{Re}\Big[
\widehat{\mathbf u}^{\,*}(\mathbf k',t)\cdot
\mathsf P(\mathbf k')\cdot
\widehat{\mathbf f}_{\rm even}(\mathbf k',t)
\Big],
\label{eq:app-even-transfer}\\
T^{\rm odd}(k,t)
&=\sum_{\mathbf k'\in\mathcal S_k}
\operatorname{Re}\Big[
\widehat{\mathbf u}^{\,*}(\mathbf k',t)\cdot
\mathsf P(\mathbf k')\cdot
\widehat{\mathbf f}_{\rm odd}(\mathbf k',t)
\Big].
\label{eq:app-odd-transfer}
\end{align}
The viscous dissipation spectrum is
\begin{equation}
D_\nu(k,t)=\nu
\sum_{\mathbf k'\in\mathcal S_k}
|\mathbf k'|^2
|\widehat{\mathbf u}(\mathbf k',t)|^2.
\label{eq:app-viscous-spectrum}
\end{equation}
\section{Direct Numerical Simulations}
\label{app:DNS}

We solve the coupled Cahn--Hilliard--Navier--Stokes equations using a Fourier pseudospectral method in a two-dimensional periodic square domain $\Omega=[0,2\pi]^2$, following the numerical framework of Ref.~\cite{padhan2025suppression}. To eliminate the pressure, we rewrite the incompressible momentum equation in the vorticity--streamfunction formulation,
\begin{equation}
\partial_t\omega+\mathbf u\cdot\nabla\omega
=\nu\nabla^2\omega+
\left[\nabla\times
(\mathbf f_{\rm even}+\mathbf f_{\rm odd})\right]
\cdot\hat{\mathbf e}_z,
\label{eq:sm_vorticity}
\end{equation}
where $\omega=(\nabla\times\mathbf u)\cdot\hat{\mathbf e}_z$ is the scalar vorticity. The streamfunction $\psi$ satisfies
\begin{equation}
\omega=-\nabla^2\psi,\qquad
\mathbf u=(\partial_y\psi,-\partial_x\psi),
\label{eq:sm_streamfunction}
\end{equation}
which automatically enforces $\nabla\cdot\mathbf u=0$.

The composition, vorticity, and streamfunction fields are represented by truncated Fourier series,
\begin{equation}
q(\mathbf x,t)=
\sum_{\mathbf k}
\widehat q(\mathbf k,t)
\exp(i\mathbf k\cdot\mathbf x),
\qquad q\in\{\phi,\omega,\psi\},
\label{eq:sm_fourier_expansion}
\end{equation}
where $\mathbf k=(k_x,k_y)$ is the Fourier wavevector, and the sum extends over all resolved Fourier modes. The evolution equations in Fourier space become
\begin{align}
\partial_t\widehat\omega
&=-\widehat{\mathbf u\cdot\nabla\omega}
-\nu k^2\widehat\omega
+i\left(k_x\widehat f_y-k_y\widehat f_x\right),
\label{eq:sm_omega_fourier}\\
\partial_t\widehat\phi
&=-\widehat{\mathbf u\cdot\nabla\phi}
-Mk^2\widehat\mu,
\label{eq:sm_phi_fourier}
\end{align}
where $k=|\mathbf k|$ and $\mathbf f=\mathbf f_{\rm even}+\mathbf f_{\rm odd}$. The chemical potential is
\begin{equation}
\widehat\mu
=\frac{3\sigma}{4\varepsilon}
\left(\widehat{\phi^3}-\widehat\phi\right)
+\frac{3\sigma\varepsilon}{2}k^2\widehat\phi.
\label{eq:sm_mu_fourier}
\end{equation}
The velocity is reconstructed from $\widehat\psi=\widehat\omega/k^2$ for $k\neq0$, with the zero mode of the streamfunction set to zero. Spatial derivatives are evaluated spectrally, while nonlinear products and capillary forcing are computed in physical space and transformed back to Fourier space. The coupled equations are advanced using the second-order exponential time-differencing Runge--Kutta method (ETDRK2)~\cite{cox2002exponential}, which treats the linear dissipative operators exactly and the remaining nonlinear contributions explicitly. We implement the Fourier pseudospectral solver in CUDA C, with GPU-optimized routines for the numerical operations. All simulations are performed on NVIDIA GPUs, including the H100 architecture.

Throughout our simulations, we fix the parameters $N=1024$, $\varepsilon=0.02$, $M=\varepsilon^2/2$, $\nu=1$, and $\sigma=1$. We define the Reynolds number as
$Re=U_{\rm rms}L_c/{\nu}$, where $U_{\rm rms}=\sqrt{2\langle e(t)\rangle_t}$ and $L_c$ is the characteristic coarsening length in the statistically stationary state. For $\chi=2$, $4$, and $10$, we obtain $Re=0.20$, $0.23$, and $0.28$, respectively, confirming that the arrested states occur at low Reynolds numbers.


\bibliographystyle{apsrev4-2}
\bibliography{ref_arxiv}

@article{pomeau2006two,
  title={Two hundred years of capillarity research},
  author={Pomeau, Yves and Villermaux, Emmanuel},
  journal={Physics Today},
  volume={59},
  number={3},
  pages={39--44},
  year={2006},
  publisher={AIP Publishing}
}

@article{cates2018theories,
  title={Theories of binary fluid mixtures: from phase-separation kinetics to active emulsions},
  author={Cates, Michael E and Tjhung, Elsen},
  journal={Journal of Fluid Mechanics},
  volume={836},
  pages={P1},
  year={2018},
  publisher={Cambridge University Press}
}

@article{hyman2014liquid,
  title={Liquid-liquid phase separation in biology},
  author={Hyman, Anthony A and Weber, Christoph A and J{\"u}licher, Frank},
  journal={Annual Review of Cell and Developmental Biology},
  volume={30},
  pages={39--58},
  year={2014},
  publisher={Annual Reviews}
}

@book{gennes2004capillarity,
  title={Capillarity and wetting phenomena: drops, bubbles, pearls, waves},
  author={Gennes, Pierre-Gilles and Brochard-Wyart, Fran{\c{c}}oise and Qu{\'e}r{\'e}, David and others},
  year={2004},
  publisher={Springer}
}

@article{bonn2009wetting,
  title={Wetting and spreading},
  author={Bonn, Daniel and Eggers, Jens and Indekeu, Joseph and Meunier, Jacques and Rolley, Etienne},
  journal={Reviews of modern physics},
  volume={81},
  number={2},
  pages={739--805},
  year={2009},
  publisher={APS}
}

@article{eggers1997nonlinear,
  title={Nonlinear dynamics and breakup of free-surface flows},
  author={Eggers, Jens},
  journal={Reviews of modern physics},
  volume={69},
  number={3},
  pages={865},
  year={1997},
  publisher={APS}
}

@article{eggers1999coalescence,
  title={Coalescence of liquid drops},
  author={Eggers, Jens and Lister, John R and Stone, Howard A},
  journal={Journal of Fluid Mechanics},
  volume={401},
  pages={293--310},
  year={1999},
  publisher={Cambridge University Press}
}

@article{siggia1979late,
  title={Late stages of spinodal decomposition in binary mixtures},
  author={Siggia, Eric D},
  journal={Physical review A},
  volume={20},
  number={2},
  pages={595},
  year={1979},
  publisher={APS}
}

@book{lighthill2001waves,
  title={Waves in fluids},
  author={Lighthill, Michael J and Lighthill, James},
  year={2001},
  publisher={Cambridge university press}
}

@article{padhan2025cahn,
  title={The Cahn--Hilliard--Navier--Stokes framework for multiphase fluid flows: laminar, turbulent and active},
  author={Padhan, Nadia Bihari and Pandit, Rahul},
  journal={Journal of Fluid Mechanics},
  volume={1010},
  pages={P1},
  year={2025},
  publisher={Cambridge University Press}
}

@article{jacqmin1999calculation,
  title={Calculation of two-phase Navier--Stokes flows using phase-field modeling},
  author={Jacqmin, David},
  journal={Journal of computational physics},
  volume={155},
  number={1},
  pages={96--127},
  year={1999},
  publisher={Elsevier}
}

@article{banerjee2017odd,
  title={Odd viscosity in chiral active fluids},
  author={Banerjee, Debarghya and Souslov, Anton and Abanov, Alexander G and Vitelli, Vincenzo},
  journal={Nature communications},
  volume={8},
  number={1},
  pages={1573},
  year={2017},
  publisher={Nature Publishing Group UK London}
}

@article{avron1998odd,
  title={Odd viscosity},
  author={Avron, JE},
  journal={Journal of statistical physics},
  volume={92},
  number={3},
  pages={543--557},
  year={1998},
  publisher={Springer}
}

@article{fruchart2023odd,
  title={Odd viscosity and odd elasticity},
  author={Fruchart, Michel and Scheibner, Colin and Vitelli, Vincenzo},
  journal={Annual Review of Condensed Matter Physics},
  volume={14},
  number={1},
  pages={471--510},
  year={2023},
  publisher={Annual Reviews}
}

@article{scheibner2020odd,
  title={Odd elasticity},
  author={Scheibner, Colin and Souslov, Anton and Banerjee, Debarghya and Sur{\'o}wka, Piotr and Irvine, William TM and Vitelli, Vincenzo},
  journal={Nature Physics},
  volume={16},
  number={4},
  pages={475--480},
  year={2020},
  publisher={Nature Publishing Group UK London}
}

@article{braverman2021topological,
  title={Topological defects in solids with odd elasticity},
  author={Braverman, Lara and Scheibner, Colin and VanSaders, Bryan and Vitelli, Vincenzo},
  journal={Physical review letters},
  volume={127},
  number={26},
  pages={268001},
  year={2021},
  publisher={APS}
}

@article{soni2019odd,
  title={The odd free surface flows of a colloidal chiral fluid},
  author={Soni, Vishal and Bililign, Ephraim S and Magkiriadou, Sofia and Sacanna, Stefano and Bartolo, Denis and Shelley, Michael J and Irvine, William TM},
  journal={Nature physics},
  volume={15},
  number={11},
  pages={1188--1194},
  year={2019},
  publisher={Nature Publishing Group UK London}
}

@article{hohenberg1977theory,
  title={Theory of dynamic critical phenomena},
  author={Hohenberg, Pierre C and Halperin, Bertrand I},
  journal={Reviews of Modern Physics},
  volume={49},
  number={3},
  pages={435},
  year={1977},
  publisher={APS}
}

@article{Tiribocchi2015,
  title={Active model H: scalar active matter in a momentum-conserving fluid},
  author={Tiribocchi, Adriano and Wittkowski, Raphael and Marenduzzo, Davide and Cates, Michael E},
  journal={Physical review letters},
  volume={115},
  number={18},
  pages={188302},
  year={2015},
  publisher={APS}
}

@article{langford2025phase,
  title={Phase separation, capillarity, and odd-surface flows in chiral active matter},
  author={Langford, Luke and Omar, Ahmad K},
  journal={Physical Review Letters},
  volume={134},
  number={6},
  pages={068301},
  year={2025},
  publisher={APS}
}

@article{wang2026edge,
  title={Edge Currents Shape Condensates in Chiral Active Matter},
  author={Wang, Boyi and Pietzonka, Patrick and J{\"u}licher, Frank},
  journal={arXiv preprint arXiv:2603.20064},
  year={2026}
}

@article{blom2026mean,
  title={Mean-Field Theory of Chiral Active Model B: Arrested Coarsening and Chiral Fingering Instabilities},
  author={Blom, Kristian and Thiele, Uwe},
  journal={arXiv preprint arXiv:2607.27305},
  year={2026}
}

@article{perlekar2017two,
  title={Two-dimensional turbulence in symmetric binary-fluid mixtures: Coarsening arrest by the inverse cascade},
  author={Perlekar, Prasad and Pal, Nairita and Pandit, Rahul},
  journal={Scientific Reports},
  volume={7},
  number={1},
  pages={44589},
  year={2017},
  publisher={Nature Publishing Group UK, London}
}

@article{perlekar2014spinodal,
  title = {Spinodal Decomposition in Homogeneous and Isotropic Turbulence},
  author = {Perlekar, Prasad and Benzi, Roberto and Clercx, Herman J. H. and Nelson, David R. and Toschi, Federico},
  journal = {Phys. Rev. Lett.},
  volume = {112},
  issue = {1},
  pages = {014502},
  numpages = {5},
  year = {2014},
  month = {Jan},
  publisher = {American Physical Society}
}

@article{active_model_H,
  title = {Active Model H: Scalar Active Matter in a Momentum-Conserving Fluid},
  author = {Tiribocchi, Adriano and Wittkowski, Raphael and Marenduzzo, Davide and Cates, Michael E.},
  journal = {Phys. Rev. Lett.},
  volume = {115},
  issue = {18},
  pages = {188302},
  numpages = {5},
  year = {2015},
  month = {Oct},
  publisher = {American Physical Society}
}

@article{padhan2024novel,
  title={Novel turbulence and coarsening arrest in active-scalar fluids},
  author={Padhan, Nadia Bihari and Kiran, Kolluru Venkata and Pandit, Rahul},
  journal={Soft Matter},
  volume={20},
  number={17},
  pages={3620--3627},
  year={2024},
  publisher={The Royal Society of Chemistry}
}

@article{singh2019droplet,
  title = {Hydrodynamically Interrupted Droplet Growth in Scalar Active Matter},
  author = {Singh, Rajesh and Cates, M. E.},
  journal = {Phys. Rev. Lett.},
  volume = {123},
  issue = {14},
  pages = {148005},
  numpages = {6},
  year = {2019},
  month = {Oct},
  publisher = {American Physical Society}
}

@article{abanov2018odd,
	title = {Odd surface waves in two-dimensional incompressible fluids},
	pages = {010},
	author = {Abanov, Alexander G. and Can, Tankut and Ganeshan, Sriram},
	journal = {SciPost Phys.},
	year = {2018},
	publisher = {SciPost}
}

@article{hosaka2026odd,
  title={Odd slip at chiral active surfaces},
  author={Hosaka, Yuto and Vilfan, Andrej},
  journal={arXiv preprint arXiv:2609.17005},
  year={2026}
}

@article{maire2026kinetic,
  title={Kinetic theory of chiral active disks: Odd transport and torque density},
  author={Maire, Rapha{\"e}l and Petrini, Alessandro and Marini Bettolo Marconi, Umberto and Caprini, Lorenzo},
  journal={The Journal of Chemical Physics},
  volume={165},
  number={3},
  year={2026},
  publisher={AIP Publishing}
}

@article{deng2023interfacial,
  title={Interfacial flow around a pusher bacterium},
  author={Deng, Jiayi and Molaei, Mehdi and Chisholm, Nicholas G and Stebe, Kathleen J},
  journal={Journal of Fluid Mechanics},
  volume={976},
  pages={A18},
  year={2023},
  publisher={Cambridge University Press}
}

@article{deng2020motile,
  title={Motile bacteria at oil--water interfaces: Pseudomonas aeruginosa},
  author={Deng, Jiayi and Molaei, Mehdi and Chisholm, Nicholas G and Stebe, Kathleen J},
  journal={Langmuir},
  volume={36},
  number={25},
  pages={6888--6902},
  year={2020},
  publisher={ACS Publications}
}

@article{cox2002exponential,
  title={Exponential time differencing for stiff systems},
  author={Cox, Steven M and Matthews, Paul C},
  journal={Journal of Computational Physics},
  volume={176},
  number={2},
  pages={430--455},
  year={2002},
  publisher={Elsevier}
}

@article{dewangan2019rotating,
  title={Rotating oil droplets driven by motile bacteria at interfaces},
  author={Dewangan, Narendra K and Conrad, Jacinta C},
  journal={Soft Matter},
  volume={15},
  number={45},
  pages={9368--9375},
  year={2019},
  publisher={The Royal Society of Chemistry}
}

@article{barker2023fluctuating,
  title={Fluctuating hydrodynamics and the Rayleigh--Plateau instability},
  author={Barker, Bryn and Bell, John B and Garcia, Alejandro L},
  journal={Proceedings of the National Academy of Sciences},
  volume={120},
  number={30},
  pages={e2306088120},
  year={2023},
  publisher={National Academy of Sciences}
}

@article{franca2026odd,
  title = {Odd droplets: Fluids with odd viscosity and highly deformable interfaces},
  author = {Fran{\ifmmode \mbox{\c{c}}\else \c{c}\fi{}}a, Hugo and Jalaal, Maziyar},
  journal = {Phys. Rev. Res.},
  volume = {8},
  issue = {3},
  pages = {033140},
  numpages = {16},
  year = {2026},
  month = {Aug},
  publisher = {American Physical Society}
}

@article{kole2021active,
  title = {Layered Chiral Active Matter: Beyond Odd Elasticity},
  author = {Kole, S. J. and Alexander, Gareth P. and Ramaswamy, Sriram and Maitra, Ananyo},
  journal = {Phys. Rev. Lett.},
  volume = {126},
  issue = {24},
  pages = {248001},
  numpages = {7},
  year = {2021},
  month = {Jun},
  publisher = {American Physical Society}
}

@article{burekovic2026bulk,
  title={Bulk and microphase separation in chiral active systems},
  author={Burekovi{\'c}, Sumeja and Kole, SJ and Maitra, Ananyo and Nardini, Cesare},
  journal={arXiv preprint arXiv:2607.12045},
  year={2026}
}

@article{torquato2018hyperuniform,
  title={Hyperuniform states of matter},
  author={Torquato, Salvatore},
  journal={Physics Reports},
  volume={745},
  pages={1--95},
  year={2018},
  publisher={Elsevier}
}

@article{padhan2025suppression,
  title={Suppression of hyperuniformity in hydrodynamic scalar active field theories},
  author={Padhan, Nadia Bihari and Voigt, Axel},
  journal={Journal of Physics: Condensed Matter},
  volume={37},
  number={10},
  pages={105101},
  year={2025},
  publisher={IOP Publishing}
}

@article{backofen2026self,
  title={Self-organized flows break morphological symmetry in active/passive systems},
  author={Backofen, Rainer and Voigt, Axel},
  journal={arXiv preprint arXiv:2601.15001},
  year={2026}
}

@article{backofen2024nonequilibrium,
  title={Nonequilibrium hyperuniform states in active turbulence},
  author={Backofen, Rainer and Altawil, Abdelrahman YA and Salvalaglio, Marco and Voigt, Axel},
  journal={Proceedings of the National Academy of Sciences},
  volume={121},
  number={24},
  pages={e2320719121},
  year={2024},
  publisher={National Academy of Sciences}
}

@article{maji2026non,
  title={Non-reciprocal Binary-fluid Turbulence},
  author={Maji, Biswajit and Padhan, Nadia Bihari and Voigt, Axel and Pandit, Rahul},
  journal={arXiv preprint arXiv:2602.06928},
  year={2026}
}

@article{perlekar2019kinetic,
  title={Kinetic energy spectra and flux in turbulent phase-separating symmetric binary-fluid mixtures},
  author={Perlekar, Prasad},
  journal={Journal of Fluid Mechanics},
  volume={873},
  pages={459--474},
  year={2019},
  publisher={Cambridge University Press}
}

@article{alexakis2018cascades,
  title={Cascades and transitions in turbulent flows},
  author={Alexakis, Alexandros and Biferale, Luca},
  journal={Physics Reports},
  volume={767},
  pages={1--101},
  year={2018},
  publisher={Elsevier}
}

@book{verma2019energy,
  title={Energy transfers in fluid flows: multiscale and spectral perspectives},
  author={Verma, Mahendra K},
  year={2019},
  publisher={Cambridge University Press}
}

@article{padhan2024interface,
  title = {Interface-induced turbulence in viscous binary fluid mixtures},
  author = {Padhan, Nadia Bihari and Vincenzi, Dario and Pandit, Rahul},
  journal = {Phys. Rev. Fluids},
  volume = {9},
  issue = {12},
  pages = {L122401},
  numpages = {9},
  year = {2024},
  month = {Dec},
  publisher = {American Physical Society},
}

@article{marchetti2022suppressed,
  title = {Activity-Suppressed Phase Separation},
  author = {Caballero, Fernando and Marchetti, M. Cristina},
  journal = {Phys. Rev. Lett.},
  volume = {129},
  issue = {26},
  pages = {268002},
  numpages = {6},
  year = {2022},
  month = {Dec},
  publisher = {American Physical Society}
}

@article{berti2005arrest,
  title = {Turbulence and Coarsening in Active and Passive Binary Mixtures},
  author = {Berti, S. and Boffetta, G. and Cencini, M. and Vulpiani, A.},
  journal = {Phys. Rev. Lett.},
  volume = {95},
  issue = {22},
  pages = {224501},
  numpages = {4},
  year = {2005},
  month = {Nov},
  publisher = {American Physical Society}
}

@article{maji2025emergent,
  title={Emergent turbulence and coarsening arrest in active-spinner fluids},
  author={Maji, Biswajit and Padhan, Nadia Bihari and Pandit, Rahul},
  journal={Communications Physics},
  volume={8},
  number={1},
  pages={488},
  year={2025},
  publisher={Nature Publishing Group UK, London},
}

@article{abels2009diffuse,
  title={On a diffuse interface model for two-phase flows of viscous, incompressible fluids with matched densities},
  author={Abels, Helmut},
  journal={Archive for rational mechanics and analysis},
  volume={194},
  number={2},
  pages={463--506},
  year={2009},
  publisher={Springer}
}

\end{document}